# Quantum coherence in molecular dynamics induced by inelastic free electron scattering

Akshay Kumar, Suvasis Swain, and Vaibhav S. Prabhudesai*

*Tata Institute of Fundamental Research, Colaba Mumbai 400005 India*

*vaibhav@tifr.res.in

**Abstract:**

**Photon-based chemical control uses photoabsorption-induced quantum coherence. Inversion symmetry breaking in the photodissociation of $H_2$ is its prime example. On the other hand, attachment of a single free electron from an incoherent source, resulting in dissociation of molecular anion formed, has also shown such symmetry-breaking. Simultaneous transfer of multiple angular momenta during the electron attachment induces the required coherence in the system by preparing the coherent superposition of two or more anion states. The even and odd values of angular momentum quanta transferred to the molecule provide the required interfering quantum paths with opposite parity. However, both these processes are resonant and occur for projectiles with a specific energy. Here we show, for the first time, such inversion symmetry-breaking in the most general scenario of non-resonant inelastic electron scattering by $H_2$, resulting in its dissociation. The ion-pair formation ($H^+ + H^-$) that proceeds after the electron impact excitation of $H_2$ shows the forward-backward asymmetry in the $H^-$ ejection about the incoming electron beam. Due to competition with autoionization, this phenomenon also shows an isotope effect. The non-resonant nature of this process makes this effect generic and points to the possible prevalent role of the underlying coherence in electron-induced chemistry.**

In modern scientific endeavors, controlling chemical reactions has been one of the most sought-after goals. In this context, coherent control of molecular dynamics using light has been explored quite extensively since the invention of lasers. The coherence induced due to the absorption of coherent photons forms the basis of the mechanism behind this control [1]. In this context, single electron attachment to a molecule also induces coherence in the molecular dynamics [2]. Here, the attachment of a free electron causes the transfer of multiple quanta of angular momentum to the molecule. These momentum transfer channels that originate due to electron attachment are inherently coherent, introducing coherence in the dissociation dynamics in the produced anion. Although coherence is induced in every such attachment process, its undisputed signature is observed in the homonuclear diatomic molecule. Such multiple momentum transfer channels that individually have inversion symmetry may invoke the formation of coherent superposition of the multiple available anion resonance states. As these individual anion states are associated with transferring either an even or odd number of angular momentum quantum to the ground molecular state, they show inversion symmetric or antisymmetric characteristics. However, their coherent superposition would not possess the inversion symmetry. Subsequent to the electron capture, the dissociation process shows the breaking of inversion symmetry. It appears in terms of forward-backward asymmetry in the fragment anion ejection with respect to the incoming electron beam. This process is analogous to the interference of two quantum paths invoked by the absorption of one or two photons. However, both photoabsorption and electron attachment are resonant processes. They occur at a specific photon or electron energy at which such excited neutral or anion states are present.

A more common feature of free electron molecule collision is non-resonant inelastic scattering, in which the incoming electron excites the target molecule by transferring part of its kinetic energy. Inelastic electron collision with molecules takes place wherever free electrons are available. Electron attachment is a subset of various phenomena that occur due to this inelastic scattering. The most extensively studied phenomenon is electron impact ionization. However, electron impact excitation and dissociation are equally important but relatively less explored channels. Except for the electron attachment, all other processes are non-resonant processes where the channel opens up if the incoming electron has energy more than its threshold energy. For example, the ionization potential for the $H_2$ molecule is 15.4 eV, and the inelastic scattering of an electron with energy more than that may cause the ionization of the target. These inelastic scattering channels are significant in plasma processes, radiation biology, astrochemistry, and any other environment that involves the electrons that are energetic enough to trigger these processes. They are responsible for the chemistry of such environments and their characteristics. However, each such collision would associate with angular momentum transfer to the molecule, similar to what is observed in the electron attachment process. Then would they also invoke the coherent dynamics in the molecule?

Ion-pair production, also known as dipolar dissociation, results in charge-asymmetric dissociation of the excited molecule. In $H_2$, this process would result in H+ + H− formation. The minimum energy required for obtaining this channel would be about 17.3 eV [3]. This process is understood as a predissociation of the excited $H_2$ molecule via the ion-pair potential energy curve of the system, as shown in Fig. 1. Hence, inelastic scattering of free electrons with an energy greater than 17.3 eV would show this channel. The ion yield curve of the H− ions from $H_2$ shows a steady increase in the ion signal from 17.3 eV as this is a non-resonant process. The incoming electron with sufficient energy excites the molecular hydrogen to the state above this threshold energy. On survival against the autoionization, this state may predissociate on the ion-pair curve to the ion-pair formation limit. Energetically, this limit lies above the neutral $H_2$ dissociation limit of H(1s) + H(n=2, 3, and 4) (Fig 1). Hence, only those curves dissociating to these limits would cross the ion-pair potential energy curve. Moreover, due to angular momentum consideration, the ion-pair formation limit $H^+$ + $H^-$ ($^1S$) corresponds to only $^2\Sigma_u^+$ and $^2\Sigma_g^+$ states. Hence, the excited target states of only these symmetries and dissociating to the H(1s) + H(n=2, 3, and 4) would contribute to the ion-pair formation near the threshold. These states belong to the series of states converging to the ground state of the $H_2^+$ cation, and they need to survive against the autoionization to the ground cation state to contribute to the ion-pair formation signal. The $H^-$ from these states will appear with very little kinetic energy, as seen from the Franck-Condon region of the $H_2$ molecule potential energy curves [4]. Being a threshold process, the kinetic energy of the $H^-$ ions produced would not change with the electron energy.

We obtained the velocity slice image (VSI) of $H^-$ ions as a function of electron energy starting from the threshold of the ion-pair formation. The experimental details are described in Methods. Starting from the 17.3 eV electron energy, the image shows a blob, as expected from the threshold process. As the electron energy increases, the blob size does not change, which is consistent with the non-resonant process where the scattered electron carries the remaining energy of the system.

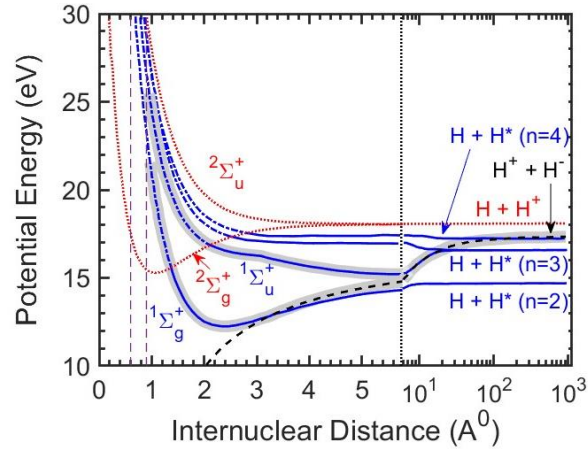

Fig 1: Schematic potential energy curves for $H_2$ relevant to forming higher kinetic energy $H^-$ ions through ion-pair formation (taken from Ref. 4). The potential energy scale is about the v=0 state of the electronic ground state of $H_2$. The red dotted curves are the $H_2^+$ ground and first excited states. The black dashed curve is the ion-pair curve, and the blue curves are the neutral excited states from the $Q_1$ series. The blue curves are partly shown as dashed-dot curves indicating that the autoionization is active in this region. The gray thick shaded curves indicate the paths taken by the $^1\Sigma_g^+$ and $^1\Sigma_u^+$ states after excitation by the incoming electrons leading to the ion-pair formation (see the text). Vertical purple dashed lines indicate the Franck-Condon region with the vibrational ground level of the electronic ground state of $H_2$.

From upwards of 25 eV electron energy, a structure in the form of a ring begins to appear in the VSI image, indicating the formation of ions with higher kinetic energy. This observed ring widens in its outer radius until the electron energy is 40 eV, and the ring size remains unchanged beyond it. Fig. 2(a) and (b) show the VSIs obtained for $H_2$ and $D_2$, respectively, at 50 eV electron energy.

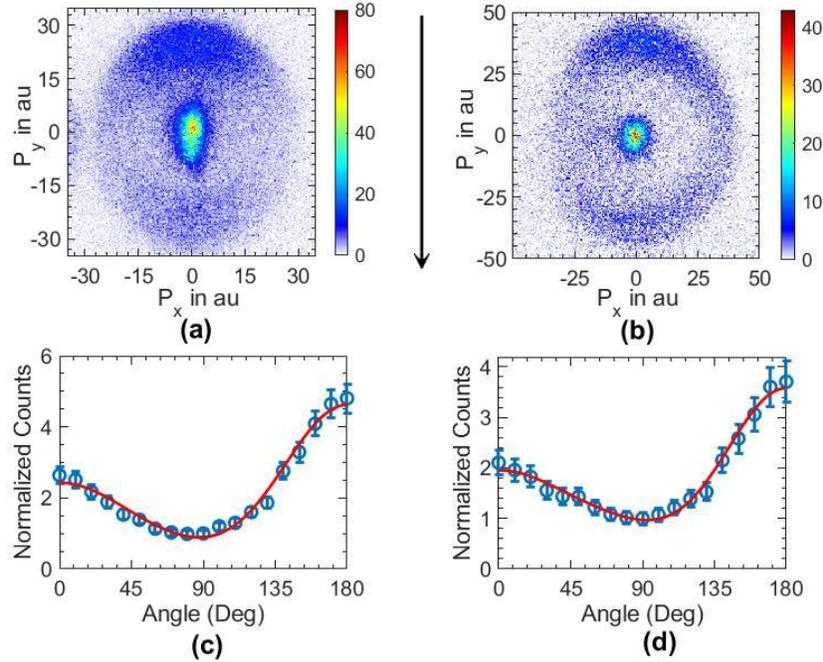

Fig. 2: Velocity slice image of (a) $H^-$ from $H_2$ and (b) $D^-$ from $D_2$ obtained from the ion-pair formation at the 50 eV electron energy. The arrow indicates the direction of the incoming electron beam. (c) and (d) are the angular distributions obtained for the kinetic energy range of 2.5 to 9 eV from the two images (a) and (b), respectively. The angular distributions are normalized w.r.t. the counts at $90^0$. The solid curves in (c) and (d) are fits for the data obtained using equation 4, discussed in the text below.

The central blob in these images corresponds to the ion-pair signal that starts appearing from the threshold, whereas the outer ring corresponds to the higher kinetic energy channel of the ion-pair formation. Fig. 2(c) and (d) show the corresponding kinetic energy-integrated (2.5 eV to 9 eV) angular distribution obtained with respect to the incoming electron from the larger energy ring. We identify this ring as the second channel contributing to ion-pair formation. The neutral excited states of $H_2$ belonging to the Rydberg series ($Q_1$ series) converging to the first excited state of $H_2^+$ contribute to this channel [5]. Similar to the excited state from the first Rydberg series, only $^1\Sigma_g^+$ and $^1\Sigma_u^+$ states from the $Q_1$ series dissociating to the H(1s) + H(n=2, 3, and 4) limit would contribute to this signal. The angular distribution of this channel shows the peak in the forward and backward directions with a non-zero signal at all angles. Interestingly, the angular distribution shows forward-backward asymmetry with more intensity in the backward direction. This observation is intriguing as it is not expected from the homonuclear diatomic molecule due to its inherent inversion symmetry, which must show up as a forward-backward symmetric dissociation pattern. The momentum image obtained for the 10 eV dissociative attachment (DA) resonance from $H_2$ shows no such asymmetry ruling out any instrumental artifact [6]. The overall signal strength is poorer in $D_2$ than in $H_2$ due to the isotope effect, where the heavier isotope experiences more loss in the population due to autoionization, as pointed out by Krishnakumar *et al.* [7]. We have measured the forward-backward asymmetry in terms of the asymmetry parameter $\eta$ defined as

$$\eta = \frac{I_F - I_B}{I_F + I_B} \qquad (1)$$

where $I_F$ and $I_B$ are the intensities of signal in the forward and backward direction with respect to the electron beam integrated over the kinetic energy range of 2.5 to 9 eV. Fig. 3 shows the asymmetry parameter, $\eta$, obtained experimentally as a function of the incoming electron energy. It also shows the asymmetry parameter obtained for $D_2$ at 50 eV electron energy.

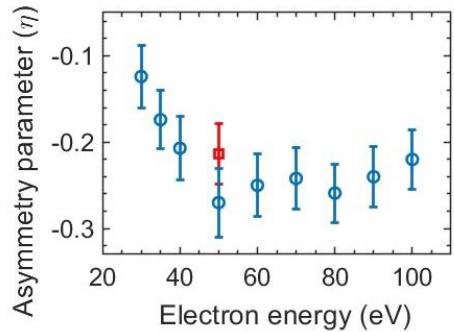

Fig. 3: Experimentally obtained kinetic energy-integrated asymmetry parameters for ion-pair formation from $H_2$ as a function of electron energy (blue circles). Also shown is the asymmetry parameter obtained for $D_2$ at 50 eV electron energy (red square).

In the inelastic scattering process, the scattered electron induces the recoil motion in the molecular target, which influences the momentum distribution of the fragments in the laboratory frame. Such an effect is expected to manifest as the distortion in the angular distribution measured by the spectrometer that detects the ions in a narrow kinetic energy range [8]. Such a spectrometer has the energy-dependent ion acceptance influencing the measurements. We have used the VSI technique to obtain the momentum image, which provides the angular distribution for all the kinetic energies in one go. We observe the forward-backward asymmetry for the signal integrated over the entire kinetic energy range, eliminating any such artifact due to molecular recoil.

The angular distribution of the molecular dissociation on electron impact is discussed in detail by Van Brunt [9]. The angular distribution at the threshold, i.e., entire electron energy transferred to the excitation of the diatomic molecule, is expected to follow that of the DA process. However, as the electron energy increases and the scattered electron carries away the excess energy, the angular distribution is expected to show changes with electron energy. It is predominantly due to the role of higher partial waves in the excitation process. However, as the initial and the final target states have specific inversion symmetry, the dissociation process is expected to retain this overall inversion symmetry. It implies that only odd or even partial waves play a role in the excitation process for a homonuclear diatomic molecule depending on the target's initial and final state. For example, in the case of H$_2$, the $^1\Sigma_g^+ \rightarrow {}^1\Sigma_g^+$ transition will have contributions only from the even partial waves, and that for the $^1\Sigma_g^+ \rightarrow {}^1\Sigma_u^+$ transition would be only odd partial waves. As a result, the angular distribution would always be symmetric about $90^0$ to the incoming electron beam if the dissociation involves only a single state excitation. Van Brunt provided the mathematical formulation to estimate the angular distribution of the ion-pair formation process [9]. At the threshold energy, the angular distribution is given by

$$I(\theta, K) = K^{-n} \left| \sum_{l=[\mu]}^{\infty} i^l i^{m(l+1)} \sqrt{\frac{(2l+1)(l-\mu)!}{(l+\mu)!}} j_l(Kr) Y_{l,\mu}(\theta, \phi) \right|^2 \qquad (2)$$

where $n$, $m$, and $r$ are adjustable parameters, $K$ is the magnitude of momentum transfer from electron to molecule, $j_l(Kr)$ is the spherical Bessel function, and $Y_{l,\mu}(\theta, \phi)$ is the spherical harmonics. For the non-threshold values of the incoming electron energies, the effective angular distribution can be estimated by integrating the $I(\theta. K)$ as

$$I(\theta) = \int_{K_0 - K_f}^{K_0 + K_f} I(\theta, K) K dK \qquad (3)$$

where $K_0$ is the wavenumber of the incoming electron and $K_f$ is the final wavenumber of the scattered electron. We have taken $n = 6$, although the final function does not vary much with this [9]. For the $^1\Sigma_g^+ \rightarrow {}^1\Sigma_u^+$ transition, we have taken m = 1, and for the $^1\Sigma_g^+ \rightarrow {}^1\Sigma_g^+$ transition, we take m = 0 [9]. The lower partial waves play a significant role close to the threshold, i.e., electron energy close to the excitation energy of the molecular state. However, as the electron energy increases, higher partial waves may become dominant with a higher value of $Kr$, increasing their weights. As the observed angular distribution does not show any sharp structures, we restrict our angular distribution functions to the lower allowed partial waves for each transition.

Based on the above equations, the observed forward-backward asymmetry in the angular distribution cannot be explained using just one of these states. We describe the physics behind this process below by using coherent superposition of the two states of opposite parities, which will have coherent contributions from both, even and odd partial waves. After considering the $s$ and $d$-wave contribution in the $^1\Sigma_g^+ \rightarrow {}^1\Sigma_g^+$ transition and *p*-wave contribution to the $^1\Sigma_g^+ \rightarrow {}^1\Sigma_u^+$ transition, we obtain the fitting function for the observed angular distribution as

$$I(\theta) = \int_{K_0-K_f}^{K_0+K_f} K^{-6} |a(j_0(Kr)Y_{0,0}(\theta,\varphi) + j_2(Kr)\sqrt{5}Y_{2,0}(\theta,\varphi)e^{-i\phi_1}) + b\sqrt{3}j_1(Kr)Y_{1,0}(\theta,\varphi)e^{-i\phi_2}|^2 KdK \quad (4)$$

The obtained fit for the data is shown in Fig. 2(c) for H$_2$ and 2(d) for D$_2$ at 50 eV electron energy and *r*=1.5 au.

As mentioned above, the states involved in this process belong to the *Q$_1$* series of autoionizing states, which are repulsive in the Franck-Condon region [5]. Due to predissociation on the ion-pair potential energy curve, these states would contribute to the measured H⁻ channel. The Franck-Condon overlap with these states would extend from 25 eV to 40 eV, making them dissociate with the kinetic energies observed in the outer ring of the VSI obtained in our experiment [4]. Guberman calculated the potential energy curves for these states [10]. Sanchez and Martin have also calculated these curves and their width towards autoionization [11]. We have used these values and the composite potential energy curve interpolated from the curves compiled by Vogel [4]. We provide below the model to explain the observed asymmetry in the angular distribution.

The ion-pair formation process is analogous to the DA process in the sense that both the processes compete with the decay of the underlying states by electron ejection. For DA, the parent negative ion resonance state may decay by autodetachment of the extra electron, whereas in ion-pair formation, the neutral excited state may decay by autoionization. However, the most crucial difference between the two processes is that DA is a resonant process that involves a molecular negative ion state. In contrast, the ion-pair formation is a non-resonant process and would occur at electron energies above its threshold. In this model, we consider only the lowest $^1\Sigma_g^+$ and $^1\Sigma_u^+$ states in the Franck-Condon region from the *Q$_1$* series (Fig. 1). We estimate the initial population of each of the states as a function of the transferred energy proportional to the corresponding part of the ground state vibrational wavefunction in the Franck-Condon region. Using autoionization width as a function of internuclear separation from Ref. 11, we determine the amplitude of each state that would survive autoionization. Subsequently, this wavepacket would dissociate along the ion-pair curve due to curve crossing. We estimate the effective amplitude of the dissociating wavepackets along the ion-pair curve by multiplying the survived wavepacket amplitude by the Landau-Zener factor that gives the transition probability between the diabatic states as [12]

$$P_{ij} = 1 - \exp\left(\frac{-2\pi c_{ij}^2}{\hbar \alpha v}\right) \quad (5)$$

where $c_{ij}$ is the coupling matrix element between the crossing states $i$ and $j$. It effectively equals half of the closest energy separation between the two relevant adiabatic potential energy curves. The typical half-splitting value for $H_2$ is about 0.27 eV [13]. The absolute slope difference (at the crossing) between the adiabatic potentials is denoted as $\alpha$, while $v$ is the relative velocity of the fragments. We estimate the phase difference between the two paths along the two $Q_1$ potential energy curves that cross the ion-pair curve at higher internuclear distances (gray shaded curves in Fig 1). For both the paths, we have used the same ion-pair curve. Using this phase difference and the remaining amplitudes of the two channels, we estimate the expected forward-backward asymmetry from the expected angular distribution function

$$I(\theta) = \int_{K_0-K_f}^{K_0+K_f} K^{-6} |a_g(j_0(Kr)Y_{0,0}(\theta,\varphi) + j_2(Kr)\sqrt{5}Y_{2,0}(\theta,\varphi)e^{-i\phi_1}) + a_u\sqrt{3}j_1(Kr)Y_{1,0}(\theta,\varphi)e^{-i\phi_2}|^2 KdK \quad (6)$$

where the $a_g$ and $a_u$ are the leftover amplitude in each channel after the loss due to autoionization and appropriate transition to the corresponding ion-pair states. We take $\phi_1$ as the initial phase difference between the $s$ and $d$-waves, which is $\pi$. $\phi_2$ is the sum of the relative phase gained during the dissociation along the two paths and the initial relative phase between the $s$ and $p$-waves, which is $\pi/2$. The asymmetry parameter $\eta$ is determined according to equation (1). The $I_F$ and $I_B$ are obtained by adding the integrated value of $I(\theta)$ from equation (6) over 0 to $\pi/2$ and $\pi/2$ to $\pi$, respectively, for each value of $K_f$. Here, parameter r, equivalent to the impact parameter, does not have a well-defined value. We estimate the asymmetry parameter over a range of values for r. The estimated asymmetry parameter as a function of incident electron energy and the impact parameter range is shown in Fig. 4(a) for $H_2$.

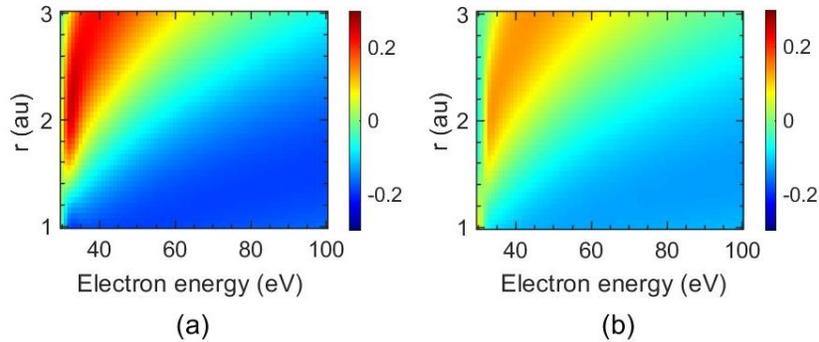

Fig. 4: The asymmetry parameter ($\eta$) as false color plots obtained from the model for (a) $H_2$ and (b) $D_2$ as a function of incoming electron energy for various values of the '$r$' parameter.

We also show the estimated values for $D_2$, which shows the isotope effect (Fig. 4(b)). The heavier isotope would affect the process in terms of the reduced amplitudes of the interfering wavepackets and the relative phase between them. Due to steep dissociation, which results in higher kinetic energy, we do not see a substantial difference between the two isotopes. However, the asymmetry diminishes for the heavier isotope as the width of the participating $^1\Sigma_g^+$ state is almost 50% larger than the $^1\Sigma_u^+$ state affecting the contrast of the interference [11]. The values obtained by the model are lower than the measured asymmetry parameters (Fig. 3). However, the trend in the observed asymmetry parameters can easily be observed in the simulated results for the typical impact parameters comparable to the equilibrium bond length of the molecule. Interestingly, such a forward-backward asymmetry was observed in ion-pair formation from electron impact with $O_2$ [14-16]. As expected such an asymmetry is not observed in two photon absorption process leading to ion-pair formation in $O_2$ as there are not multiple paths of opposite parities involved in the dissociation process [17].

To conclude, we have shown that the ion-pair formation process in $H_2$ resulting from non-resonant inelastic scattering of electrons shows inversion symmetry-breaking. The quantum coherence induced by the transfer of odd and even partial waves results in the transition of the molecule to the superposition of two states from the $Q_1$ band with opposite parity and dissociating to the same limit. The non-resonant nature of this process makes these results more generic than the DA or photodissociation processes that are resonant in nature. The above results indicate that similar asymmetry should be expected in the electron impact dissociation of inversion symmetric molecules resulting in the fragments in the ground and excited states, for example, in the electron impact dissociation of $H_2$ into $H + H^*$. These results put forward the intriguing aspect of inelastic scattering of the particles that invokes a coherent response from the matter. Although the coherence induced in the non-resonant inelastic scattering of electrons from an incoherent source is clearly observed in the homonuclear diatomic system, such coherence-induced effects must be prevalent in the general electron scattering from any molecules. Recently, low energy free electron attachment has also been shown to control molecular dissociation, which is the first step towards realizing the ultimate control in the chemical reaction [18]. In addition, there have been extensive explorations of low-energy electron-induced control over molecular dynamics, including efforts toward single-molecule engineering using a scanning tunneling microscope [19]. Our results point to yet another possible avenue of electron-induced chemical control where the neutral excited states are accessed for molecular dissociation. However, it remains to be seen how we can tap these coherences to obtain total control over chemical reactions.

**Acknowledgment:**

We thank E. Krishnakumar for the valuable discussions. All authors acknowledge the financial support from the Dept. of Atomic Energy, India, under Project Identification No. RTI4002.

**Contributions**

VSP planned the research. VSP and SS built the experiment with help from AK AK, and SS carried out the measurements. AK and VSP carried out the simulations and analysis, and VSP and AK interpreted the results. VSP prepared the manuscript.

**Methods:**

The details of the experimental setup are given elsewhere [6]. Here we provide a brief description of the experimental scheme. A pulsed (100 ns) electron beam collimated by a 50 G magnetic field is made to cross an effusive molecular beam produced by a capillary array. The resulting H¯ ions were extracted into the velocity slice imaging (VSI) spectrometer using a pulsed extraction field. We detect the ions using a two-dimensional position-sensitive detector made of a pair of microchannel plates followed by the phosphor screen. We have improved the imaging resolution over the previous experiment [6] by using a high voltage switch of 10 ns duration compared to 80 ns used earlier. This modification has enabled us to detect the relatively fast-moving H¯ ions by applying higher extraction voltages. The images recorded using a charge-coupled device camera were analyzed offline for the H¯ ions' kinetic energy and angular distribution. We carry out the electron energy calibration using the dissociative attachment (DA) signal from $H_2$ at 14 eV and the imaging calibration by measuring the VSI of H¯ ions from the 10eV DA peak of $H_2$. The images are obtained from the crossed-beam geometry of the target region by subtracting the contribution from the static gas background [6].